\documentclass[11pt]{article}
\usepackage{amsmath,amsfonts,amsthm,times,latexsym,graphicx}

\newtheorem{lemma}{Lemma}[section]
\newtheorem{corollary}[lemma]{Corollary}

\newenvironment{Lemma}
	{\begin{lemma}\sl}
	{\end{lemma}}
\newenvironment{Corollary}
	{\begin{corollary}\sl}
	{\end{corollary}}

\newcommand{\ruck}[1]{\strut\hspace{#1cm}}

\def\bs{\boldsymbol}

\def\R{\mathbb{R}}

\def\eps{\epsilon}

\def\GG{\mathcal{G}}

\def\argmax{\mathop{\rm arg\,max}}

\def\Ex{\mathop{\rm I\!E}\nolimits}
\def\Pr{\mathop{\rm I\!P}\nolimits}

\begin{document}

\addtolength{\baselineskip}{+.3\baselineskip}

\thispagestyle{empty}

\begin{center}
	{\large University of Bern}
	
	{\large Institute of Mathematical Statistics and Actuarial Science}
	
	\textbf{\large Technical Report 71}
\end{center}

\vfill

\begin{center}
	\textbf{\Large On an Auxiliary Function for Log-Density Estimation}

	{\large Madeleine L.\ Cule and Lutz D\"umbgen}
	
	\textsl{(University of Cambridge and University of Bern)}
	
	{\large July 2008, minor revisions in January 2016}
\end{center}

\vfill

\begin{abstract}
In this note we provide explicit expressions and expansions for a special function $J$ which appears in nonparametric estimation of log-densities. This function returns the integral of a log-linear function on a simplex of arbitrary dimension. In particular it is used in the \textsl{R}-package \textsl{LogCondDEAD} by Cule et al.\ (2007).
\end{abstract}

\vfill
\vfill

\strut

\newpage

\tableofcontents

\vskip 1cm

\section{Introduction}

Suppose one wants to estimate a probability density $f$ on a certain compact region $C \subset \R^d$, based on an empirical distribution $\hat{P}$ of a sample from $f$. One possibility is to embed $C$ into a union
\[
	S \ = \ \bigcup_{j=1}^m S_j
\]
of simplices $S_j \subset \R^d$ with pairwise disjoint interior. By a simplex in $\R^d$ we mean the convex hull of $d+1$ points. Then we consider the family $\GG = \GG(S_1,\ldots,S_m)$ of all continuous functions $\psi : S \to \R$ which are linear on each simplex $S_j$. Now
\begin{equation}
\label{eq: def of psihat}
	\hat{\psi} \ := \ \argmax_{\psi \in \GG} \left( \int_S \psi \, d \hat{P} - \int_S \exp(\psi(x)) \, dx \right)
\end{equation}
defines a maximum likelihood estimator $\hat{f} := \exp(\hat{\psi})$ of a probability density on $S$, based on $\hat{P}$. For existence and uniqueness of this estimator see, for instance, Cule et al.\ (2008).

To compute $\hat{\psi}$ explicitly, note that $\psi \in \GG$ is uniquely determined by its values at the corners (extremal points) of all simplices $S_j$, and $\int \psi \, d\hat{P}$ is a linear function of these values. The second integral in (\ref{eq: def of psihat}) may be represented as follows: Let $S_j$ be the convex hull of $\bs{x}_{0j}, \bs{x}_{1j}, \ldots, \bs{x}_{dj} \in \R^d$, and set $y_{ij} := \psi(\bs{x}_{ij})$. Then
\[
	\int_S \exp(\psi(x)) \, dx
	\ = \ \sum_{i=1}^m \int_{S_i} \exp(\psi(x)) \, dx
	\ = \ \sum_{i=1}^m D_j \cdot J(y_{0j}, y_{1j}, \ldots, y_{dj}) ,
\]
where
\[
	D_j \ := \ \det \bigl[ \bs{x}_{1j}-\bs{x}_{0j}, \bs{x}_{2j}-\bs{x}_{0j}, \ldots,
			\bs{x}_{dj}-\bs{x}_{0j} \bigr] ,
\]
while $J(\cdot)$ is an auxiliary function defined and analyzed subsequently.

\section{The special function $\bs{J(\cdot)}$}

\subsection{Definition of $\bs{J(\cdot)}$}

For $d \in \mathbb{N}$ let
\[
	\mathcal{T}_d
	\ := \ \Bigl\{ \bs{u} \in (0,1)^d : \sum_{i=1}^d u_i < 1 \Bigr\} .
\]
Then for $y_0, y_1, \ldots, y_d \in \R$ we define
\[
	J(y_0, y_1, \ldots, y_d)
	\ := \ \int_{\mathcal{T}_d} \exp \Bigl( (1 - u_+) y_0 + \sum_{i=1}^d u_i y_i \Bigr) \, d\bs{u}
\]
with $u_+ := \sum_{i=1}^d u_i$.

Standard considerations in connection with beta- and gamma-distributions as described in Section~\ref{sec: Gamma-Beta} reveal the following alternative representation:
\[
	J(y_0, y_1, \ldots, y_d)
	\ := \ \frac{1}{d!} \,
		\mathbb{E} \exp \Bigl( \sum_{i=0}^d B_i y_i \Bigr)
\]
with $B_i = B_{d,i} := E_i \big/ \sum_{s=0}^d E_s$ and stochastically independent, standard exponential random variables $E_0, E_1, \ldots, E_d$. This representation shows clearly that $J(\cdot)$ is symmetric in its arguments.

An often useful identity is
\begin{equation}
\label{eq: Centering}
	J(y_0, y_1, \ldots, y_d)
	\ = \ \exp(y_*) J(y_0 - y_*, y_1 - y_*, \ldots, y_d - y_*)
	\quad\text{for any} \ y_* \in \R .
\end{equation}

\subsection{A first recursion formula}

For $d = 1$ one can compute $J(y_0, y_1)$ explicitly:
\[
	J(y_0, y_1)
	\ = \ \int_0^1 \exp \bigl( (1 - u) y_0 + u y_1 \bigr) \, du
	\ = \ \begin{cases}
			\displaystyle
			\frac{\exp(y_1) - \exp(y_0)}{y_1 - y_0} & \text{if} \ y_0 \ne y_1 ,
		\\[2ex]
			\exp(y_0) & \text{if} \ y_0 = y_1 .
		\end{cases}
\]
For $d \ge 2$ one may use the following recursion formula:
\begin{equation}
\label{eq: Recursion 2}
	J(y_0, y_1, \ldots, y_d)
	\ = \ \left\{\begin{array}{cl}
			\displaystyle
			\frac{J(y_1, y_2, \ldots, y_d) - J(y_0, y_2, \ldots, y_d)}
			     {y_1 - y_0}
				& \text{if} \ y_0 \ne y_1 ,
		\\[2ex]
			\displaystyle
			\frac{\partial}{\partial y_1} J(y_1, y_2, \ldots, y_d)
				& \text{if} \ y_0 = y_1 .
		\end{array}\right.
\end{equation}
Since $J(y_0,y_1,\ldots,y_d)$ is continuous in $y_0, y_1, \ldots, y_d$, it suffices to verify \eqref{eq: Recursion 2} in case of $y_0 \ne y_1$. We may identify $\mathcal{T}_d$ with the set $\bigl\{(v,\bs{u}) : \bs{u} \in \mathcal{T}_{d-1}, v \in (0,1-u_+) \bigr\}$. Then it follows from Fubini's theorem that
\begin{align*}
	J(& y_0, y_1, \ldots, y_d) \\
	&= \ \int_{\mathcal{T}_{d-1}} \int_0^{1 - u_+}
		\exp \Bigl( (1 - u_+ - v) y_0 + v y_1 + \sum_{i=2}^d u_{i-1} y_i \Bigr)
			\, dv \, d\bs{u} \\
	&= \ \int_{\mathcal{T}_{d-1}} \Bigl(
		\frac{\exp \bigl( (1 - u_+ - v) y_0 + v y_1 + \sum_{i=2}^d u_{i-1} y_i \bigr)}
			{y_1 - y_0} \Bigr) \bigg|_{v=0}^{1 - u_+} \, d\bs{u} \\
	&= \ \int_{\mathcal{T}_{d-1}}
		\frac{\exp \bigl( (1 - u_+) y_1 + \sum_{i=2}^d u_{i-1} y_i \bigr)
			- \exp \bigl( (1 - u_+) y_0 + \sum_{i=2}^d u_{i-1} y_i \bigr)}
			{y_1 - y_0} \, d\bs{u} \\
	&= \ \frac{J(y_1,y_2,\ldots,y_d) - J(y_0,y_2,\ldots,y_d)}{y_1 - y_0} .
\end{align*}

\subsection{Another recursion formula}

It is well-known that for any integer $0 \le j < d$,
\[
	\left( \frac{E_i}{\sum_{s=0}^j E_s} \right)_{i=0}^j , \quad
	B := \frac{\sum_{i=0}^j E_i}{\sum_{s=0}^d E_s} , \quad
	\left( \frac{E_i}{\sum_{s=j+1}^d E_s} \right)_{i=j+1}^d
\]
are stochastically independent with $B \sim \mathrm{Beta}(j+1, d-j)$; see also Section~\ref{sec: Gamma-Beta}. Hence we end up with the following recursive identity:
\begin{align*}
	J(& y_0, y_1, \ldots, y_d) \\
	&= \ \frac{j! (d-j-1)!}{d!} \, \mathbb{E} \bigl(
		J(B y_0, \ldots, B y_j) J((1 - B) y_{j+1}, \ldots, (1 - B) y_d) \bigr) \\
	&= \ \int_0^1 u^{j} (1 - u)^{d-j-1}
		J(u y_0, \ldots, u y_j) J((1 - u) y_{j+1}, \ldots, (1 - u) y_d) \, du
\end{align*}
with
\[
	J(r) \ := \ \exp(r) .
\]
Here we used the well-known identity
\begin{equation}
\label{eq: Beta}
	\int (1 - u)^\ell u^m \, du \ = \ \frac{\ell! m!}{(\ell + m + 1)!}
	\quad\text{for integers} \ \ell, m \ge 0 .	
\end{equation}
Plugging in $j = d-1$ into the previous recursive equation leads to
\begin{equation}
\label{eq: Recursion 1}
	J(y_0, y_1, \ldots, y_d)
	\ = \ \int_0^1 u^{d-1} J(u y_0, \ldots, u y_{d-1}) \exp((1 - u) y_d) \, du .
\end{equation}

\section{An expansion for $\bs{J(\cdot)}$}

With $\bar{y} := (d+1)^{-1} \sum_{i=0}^d y_i$ and $z_i := y_i - \bar{y}$ one may write
\[
	J(y_0, y_1, \ldots, y_d)
	\ = \ \exp(\bar{y}) J(z_0, z_1, \ldots, z_d)
\]
by virtue of (\ref{eq: Centering}). Note that $z_+ := \sum_{i=0}^d z_i = 0$. As $\bs{z} := (z_i)_{i=0}^d \to \bs{0}$,
\begin{align*}
	d! \, J(& z_0, z_1, \ldots, z_d) \\
	&= \ 1 + \sum_{i=0}^d \Ex(B_i) z_i
		+ \frac{1}{2} \sum_{i,j=0}^d \Ex(B_i B_j) z_i z_j
		+ \frac{1}{6} \sum_{i,j,k=0}^d \Ex(B_i B_j B_k) z_i z_j z_k
		+ O(\|\bs{z}\|^4) .
\end{align*}
It follows from Lemma~\ref{lemma: Gamma-Beta} that
\[
	\Ex \Bigl( \prod_{i=0}^d B_i^{k_i} \Bigr)
	\ = \ \prod_{i=0}^d k_i! \Big/ [d+k_+]_{k_+}
	\quad\text{for integers} \ k_0, k_1, \ldots, k_d \ge 0 .
\]
In particular,
\begin{align*}
	\Ex(B_0) \
	&= \ \frac{1}{d+1} , \\
	\Ex(B_0^2) \
	&= \ \frac{2}{[d+2]_2} ,
		\qquad \Ex(B_0 B_1) \ = \ \frac{1}{[d+2]_2} , \\
	\Ex(B_0^3) \
	&= \ \frac{6}{[d+3]_3} ,
		\qquad \Ex(B_0^2 B_1) \ = \ \frac{2}{[d+3]_3} ,
		\qquad \Ex(B_0 B_1 B_2) \ = \ \frac{1}{[d+3]_3} .
\end{align*}
Consequently, $\sum_{i=0}^d \Ex(B_i) z_i = \Ex(B_0) z_+ = 0$,
\begin{align*}
	[d+2]_2 \sum_{i,j=0}^d \Ex(B_i B_j) z_i z_j \
	&= \ \sum_{i,j=0}^d \bigl( 1_{[i = j]} \cdot 2 + 1_{[i \ne j]} \bigr) z_i z_j \\
	&= \ \sum_{i,j=0}^d \bigl( 1_{[i = j]} + 1 \bigr) z_i z_j \\
	&= \ \sum_{i=0}^d z_i^2 + z_+^2 \\
	&= \ \sum_{i=0}^d z_i^2 ,
\end{align*}
and
\begin{align*}
	[d+3]_3 & \sum_{i,j,k=0}^d \Ex(B_i B_j B_k) z_i z_j z_k \\
	&= \ \sum_{i,j,k=0}^d \bigl( 1_{[i=j=k]}\cdot 6 + 1_{[\#\{i,j,k\} = 2]} \cdot 2
				+ 1_{[\#\{i,j,k\} = 3]} \bigr) z_i z_j z_k \\
	&= \ \sum_{i,j,k=0}^d \bigl( 1_{[i=j=k]}\cdot 5 + 1_{[\#\{i,j,k\} = 2]} + 1 \bigr)
		z_i z_j z_k \\
	&= \ 5 \sum_{i=0}^d z_i^3 + 3 \sum_{s,t=0}^d 1_{[s \ne t]} z_s^2 z_t + z_+^3 \\
	&= \ 5 \sum_{i=0}^d z_i^3 + 3 \sum_{s=0}^d z_s^2 z_+ - 3 \sum_{s=0}^d z_s^3 + z_+^3 \\
	&= \ 2 \sum_{i=0}^d z_i^3 .
\end{align*}
Consequently,
\begin{equation}
\label{eq: Taylor of J}
	J(y_0,y_1,\ldots,y_d)
	\ = \ \exp(\bar{y}) \Bigl( \frac{1}{d!} + \frac{1}{2 (d+2)!} \sum_{i=0}^d z_i^2
			+ \frac{1}{3 (d+3)!} \sum_{i=0}^d z_i^3
			+ O \bigl( \|\bs{z}\|^4 \bigr) \Bigr) .
\end{equation}

\section{A recursive implementation of $\bs{J(\cdot)}$
	and its partial derivatives}

By means of (\ref{eq: Recursion 2}) and the Taylor expansion (\ref{eq: Taylor of J}) one can implement the function $J(\cdot)$ in a recursive fashion. In what follows we use the abbreviation
\[
	y_{a:b} \ = \ \begin{cases}
		(y_a, \ldots, y_b) & \text{if} \ a \le b \\
		() & \text{if} \ a > b
	\end{cases}
\]

To compute $J(y_{0:d})$ we assume without loss of generality that $y_0 \le y_1 \le \cdots \le y_d$.
It follows from (\ref{eq: Recursion 2}) and symmetry of $J(\cdot)$ that
\[
	J(y_{0:d}) \ = \ \frac{J(y_{1:d}) - J(y_{0:d-1})}{y_d - y_0}
\]
if $y_0 \ne y_d$. This formula is okay numerically if $y_d - y_0$ is not too small. Otherwise one should use (\ref{eq: Taylor of J}). This leads to the the pseudo code in Table~\ref{tab: J}.

\begin{table}
\centerline{\bf\begin{tabular}{|l|} \hline
\ruck{0}	Algorithm $J \leftarrow \mbox{J}(y,d,\eps)$\\
\ruck{0}	if $y_d - y_0 < \eps$ then\\
\ruck{1}		$\bar{y} \leftarrow \sum_{i=0}^d y_i / (d+1)$\\
\ruck{1}		$z_2 \leftarrow \sum_{i=0}^d (y_i - \bar{y})^2 / 2$\\
\ruck{1}		$z_3 \leftarrow \sum_{i=0}^d (y_i - \bar{y})^3 / 3$\\
\ruck{1}		$J \leftarrow \exp(\bar{y}) \bigl( 1/d! + z_2 / (d+2)! + z_3 / (d+3)! \bigr)$\\
\ruck{0}	else\\
\ruck{1}		$J \leftarrow \bigl( \mbox{J}(y_{1:d},d-1,\eps) - \mbox{J}(y_{0:d-1},d-1,\eps) \bigr)
				/ (y_d - y_0)$ \\
\ruck{0}	end if.\\\hline
\end{tabular}}
\caption{Pseudo-code for $J(y)$ with ordered input vector $y$.}
\label{tab: J}
\end{table}

To avoid messy formulae, one can express partial derivatives of $J(\cdot)$ in terms of higher order versions of $J(\cdot)$ by means of the recursion (\ref{eq: Recursion 2}). For instance,
\begin{align*}
	\frac{\partial J(y_{0:d})}{\partial y_0} \
	&= \ \lim_{\eps \to 0} \,
		\frac{J(y_0+\eps, \, y_{1:d}) - J(y_0, \, y_{1:d})}{\eps} \\
	&= \ \lim_{\eps \to 0} \, J(y_0, y_0+\eps, \, y_{1:d}) \\
	&= \ J(y_0, y_0, y_{1:d}) .
\end{align*}
Similarly,
\begin{align*}
	\frac{\partial^2 J(y_{0:d})}{\partial y_0^2} \
	&= \ \lim_{\eps \to 0} \Bigl(
		\frac{J(y_0+\eps, \, y_{1:d}) - J(y_0, \, y_{1:d})}{\eps}
			- \frac{J(y_0, \, y_{1:d}) - J(y_0-\eps, \, y_{1:d})}{\eps}
		\Bigr) \big/ \eps \\
	&= \ 2 \, \lim_{\eps \to 0} \,
		\frac{J(y_0, y_0+\eps, \, y_{1:d}) - J(y_0, y_0-\eps, \, y_{1:d})}{2\eps} \\
	&= \ 2 \, \lim_{\eps \to 0} \, J(y_0, y_0-\eps, y_0+\eps, \, y_{1:d}) \\
	&= \ 2 \, J(y_0, y_0, y_0, \, y_{1:d}) ,
\end{align*}
while
\begin{align*}
	\frac{\partial^2 J(y_{0:d})}{\partial y_0 \partial y_1} \
	&= \ \lim_{\eps \to 0} \Bigl(
			\frac{J(y_0+\eps, y_1+\eps, \, y_{2:d}) - J(y_0, y_1+\eps, \, y_{2:d})}{\eps} \\
	& \qquad\qquad\qquad - \
		\frac{J(y_0+\eps, y_1, \, y_{2:d}) - J(y_0, y_1, \, y_{2:d})}{\eps}
		\Bigr) \big/ \eps \\
	&= \ \lim_{\eps \to 0} \,
		\frac{J(y_0, y_0+\eps, y_1+\eps, \, y_{2:d}) - J(y_0, y_0+\eps, y_1, \, y_{2:d})}
			{\eps} \\
	&= \ \lim_{\eps \to 0} \, J(y_0, y_0+\eps, y_1, y_1+\eps, \, y_{2:d}) \\
	&= \ J(y_0, y_0, y_1, y_1, \, y_{2:d}) .
\end{align*}

\section{The special cases $\bs{d = 1}$ and $\bs{d = 2}$}

For small dimension $d$ it may be worthwhile to work with non-recursive implementations of the function $J(\cdot)$. Here we collect and extend some results of D\"umbgen et al.\ (2007).

\subsection{General considerations about a bivariate function}
\label{subsec: General bivariate function}

In view of (\ref{eq: Recursion 2}) we consider an arbitrary function $f : \R \to \R$ which is infinitely often differentiable. Then
\[
	h(r,s) \
	:= \ \begin{cases}
			\displaystyle
			\frac{f(s) - f(r)}{s - r}
				& \text{if} \ s \ne r \\[2ex]
			f'(r)
				& \text{if} \ s = r
		\end{cases}
\]
defines a smooth and symmetric function $h : \R^2 \to \R$ such that
\[
	h(r,s) \ = \ f'(r) + \frac{f''(r)}{2}(s - r) + O \bigl( (s - r)^2 \bigr)
				\quad\text{as} \ s \to r .
\]
Its first partial derivatives of order one and two are given by
\begin{align*}
	\frac{\partial h(r,s)}{\partial r} \
	&= \ \begin{cases}
		\displaystyle
		\frac{f(s) - f(r) - f'(r)(s - r)}{(s - r)^2}
			& \text{if} \ s \ne r , \\[2ex]
		\displaystyle
		\frac{f''(r)}{2} + \frac{f'''(r)}{6} (s - r)
					+ O \bigl( (s - r)^2 \bigr)
				& \text{as} \ s \to r ,
		\end{cases} \\[2ex]
	\frac{\partial^2 h(r,s)}{\partial r^2} \
	&= \ \begin{cases}
		\displaystyle
		\frac{2 \bigl( f(s) - f(r) - f'(r)(s - r) \bigr) - (s - r)^2 f''(r)}
		     {(s - r)^3}
			& \text{if} \ s \ne r , \\[2ex]
		\displaystyle
		\frac{f'''(r)}{3} + \frac{f''''(r)}{12} (s - r)
				+ O \bigl( (s - r)^2 \bigr)
			& \text{as} \ s \to r ,
		\end{cases} \\[2ex]
	\frac{\partial^2 h(r,s)}{\partial r \partial s} \
	&= \ \begin{cases}
		\displaystyle
		\frac{(s - r) \bigl( f'(r) + f'(s) \bigr)
		        - 2 \bigl( f(s) - f(r) \bigr)}
		     {(s - r)^3}
			& \text{if} \ s \ne r , \\[2ex]
		\displaystyle
		\frac{f'''(r)}{6} + \frac{f''''(r)}{12} (s - r)
				+ O \bigl( (s - r)^2 \bigr)
			& \text{as} \ s \to r .
		\end{cases}
\end{align*}
The other partial derivatives of order one and two follow via symmetry considerations.

\subsection{More details for the case $\bs{d = 1}$}

Recall that
\[
	J(r, s)
	\ = \ \int_0^1 \exp \bigl( (1 - u) r + u s \bigr) \, du
	\ = \ \left\{\begin{array}{cl}
		\displaystyle
		\frac{\exp(s) - \exp(r)}{s - r} & \text{if} \ r \ne s , \\[2ex]
		\exp(r) & \text{if} \ r = s .
	\end{array}\right.
\]
This is just the function introduced by D\"umbgen, H\"usler and Rufibach (2007). Let us recall some properties and formulae for the corresponding partial derivatives
\[
	J_{a,b}(r, s)
	\ := \ \frac{\partial^{a+b}}{\partial r^a \partial s^b} \, J(r, s)
	\ = \ \int_0^1 (1 - u)^a u^b \exp((1 - u) r + u s) \, du .
\]
Note first that
\[
	J_{a,b}(r,s) \ = \ J_{b,a}(s,r) \ = \ \exp(r) J_{a,b}(0, s - r) .
\]
Thus it suffices to derive formulae for $(r, s) = (0,y)$ and $b \le a$. It follows from (\ref{eq: Beta}) that
\begin{align*}
	J_{a,0}(0, y) \
	&= \ \int_0^1 (1 - u)^a \sum_{k=0}^\infty \frac{u^k}{k!} \, y^k \, du \\
	&= \ \sum_{k = 0}^\infty \frac{1}{k!} \int_0^1 (1 - u)^a u^k \, du
		\cdot y^k \\
	&= \ \sum_{k = 0}^\infty \frac{a!}{(k+a+1)!} \, y^k \\
	&= \ \frac{a!}{y^{a+1}}
		\Bigl( \exp(y) - \sum_{\ell=0}^a \frac{y^\ell}{\ell!} \Bigr) .
\end{align*}
In particular,
\begin{align*}
	J_{1,0}(0, y) \
	&= \ \frac{\exp(y) - 1 - y}{y^{2}} \\
	&= \ \frac{1}{2} + \frac{y}{6} + \frac{y^2}{24} + \frac{y^3}{120}
		+ O(y^4)	\quad	(y \to 0) , \\[2ex]
	J_{2,0}(0, y)
	&= \ \frac{2 (\exp(y) - 1 - y - y^2/2)}{y^3} \\
	&= \ \frac{1}{3} + \frac{y}{12} + \frac{y^2}{60} + \frac{y^3}{360}
		+ O(y^4)	\quad	(y \to 0) , \\[2ex]
	J_{3,0}(0, y)
	&= \ \frac{6 (\exp(y) - 1 - y - y^2/2 - y^3/6)}{y^4} \\
	&= \ \frac{1}{4} + \frac{y}{20} + \frac{y^2}{120} + \frac{y^3}{840}
		+ O(y^4)	\quad	(y \to 0) , \\[2ex]
	J_{4,0}(0, y)
	&= \ \frac{24 (\exp(y) - 1 - y - y^2/2 - y^3/6 - y^4/24)}{y^5} \\
	&= \ \frac{1}{5} + \frac{y}{30} + \frac{y^2}{210} + \frac{y^3}{1680}
		+ O(y^4)	\quad	(y \to 0) .
\end{align*}
Another general observation is that
\begin{align*}
	J_{a,b}(r,s) \
	&= \ \int_0^1 (1 - u)^a (1 - (1-u))^b \exp((1 - u) r + u s) \, du \\
	&= \ \sum_{i=0}^b \binom{b}{i} (-1)^i J_{a+i,0}(r,s) .
\end{align*}
In particular, 
\begin{align*}
	J_{a,1}(r,s) \
	&= \ J_{a,0}(r,s) - J_{a+1,0}(r,s) , \\
	J_{a,2}(r,s)
	&= \ J_{a,0}(r,s) - 2 J_{a+1,0}(r,s) + J_{a+2,0}(r,s) .
\end{align*}
On the other hand,
\begin{align*}
	J_{a,b}(0,y) \
	&= \ \sum_{k=0}^\infty \frac{y^k}{k!} \int_0^1 (1 - u)^a u^{k+b} \, du \\
	&= \ \sum_{k=0}^\infty \frac{a! [k+b]_b}{(k+a+b+1)!} \, y^k
\end{align*}
with $[r]_0 := 1$ and $[r]_m := \prod_{i=0}^{m-1} (r - i)$ for integers $m > 0$. In particular,
\begin{align*}
	J_{1,1}(0,y) \
	&= \ \frac{\exp(y)(y - 2) + 2 + y}{y^3} \\
	&= \ \frac{1}{6} + \frac{y}{12} + \frac{y^2}{40} + \frac{y^3}{180}
		+ O(y^4)	\quad	(y \to 0) .
\end{align*}

\subsection{The case $\bs{d = 2}$}

Our recursion formula (\ref{eq: Recursion 2}) yields
\[
	J(r, s, t)
	\ = \ \left\{\begin{array}{cl}
			\displaystyle
			\frac{J(s,t) - J(r,t)}{s - r} & \text{if} \ r \ne s ,
		\\[2ex]
			J_{10}(r, t) & \text{if} \ r = s .
	\end{array}\right.
\]
Because of $J$'s symmetry we may rewrite this in terms of the order statistics $y_{(0)} \le y_{(1)} \le y_{(2)}$ of $(y_i)_{i=0}^2$ as
\[
	J(r, s, t)
	\ = \ \left\{\begin{array}{cl}
			\displaystyle
			\frac{J(y_{(1)}, y_{(2)}) - J(y_{(0)}, y_{(1)})}{y_{(2)} - y_{(0)}}
				& \text{if} \ y_{(0)} < y_{(2)} , \\[2ex]
			\displaystyle
			\frac{\exp(y_{(0)})}{2}
				& \text{if} \ y_{(0)} = y_{(2)} .
		\end{array}\right.			
\]

For fixed third argument $t$, this function $J(r,s,t)$ corresponds to $h(r,s)$ in Section~\ref{subsec: General bivariate function} with $f(x) := J(x,t)$. Thus
\[
	\frac{\partial J(r, s, t)}{\partial r}
	\ = \ \left\{\begin{array}{cl}
			\displaystyle
			\frac{J(s, t) - J(r, t) - J_{1,0}(r,t)(s - r)}
			     {(s - r)^2}
				& \text{if} \ r \ne s , \\[2ex]
			\displaystyle
			\frac{J_{2,0}(r,t)}{2} + \frac{J_{3,0}(r,t)(s - r)}{6}
					+ O \bigl( (s - r)^2 \bigr)
				& \text{as} \ s \to r .
		\end{array}\right.
\]
Moreover,
\begin{align*}
	\frac{\partial^2 J(r, s, t)}{\partial r^2} \
	&= \ \begin{cases}
		\displaystyle
		\frac{2 \bigl( J(s, t) - J(r, t)
		               - J_{1,0}(r,t)(s - r) \bigr)
		        - (s - r)^2 J_{2,0}}
		     {(s - r)^3}
			& \text{if} \ r \ne s , \\[2ex]
		\displaystyle
		\frac{J_{3,0}(r,t)}{3} + \frac{J_{4,0}(r,t)(s - r)}{12}
				+ O \bigl( (s - r)^2 \bigr)
			& \text{as} \ s \to r ,
		\end{cases} \\[2ex]
	\frac{\partial^2 J(r, s, t)}{\partial r \partial s} \
	&= \ \begin{cases}
		\displaystyle
		\frac{\bigl( J_{1,0}(r,t) + J_{1,0}(s,t) \bigr) (s - r)
		        - 2 \bigl( J(s,t) - J(r,t) \bigr)}
		     {(s - r)^3}
			& \text{if} \ r \ne s , \\[2ex]
		\displaystyle
		\frac{J_{3,0}(r,t)}{6} + \frac{J_{4,0}(r,t)(s - r)}{12}
				+ O \bigl( (s - r)^2 \bigr)
			& \text{as} \ s \to r .
		\end{cases}
\end{align*}

\section{Gamma and multivariate beta (Dirichlet) distributions}
\label{sec: Gamma-Beta}

Let $G_0, G_1, \ldots, G_m$ be stochastically independent random variables with $G_i \sim \mathrm{Gamma}(a_i)$ for certain parameters $a_i > 0$. That means, for any Borel set $A \subset (0,\infty)$,
\[
	\Pr(G_i \in A)
	\ = \ \int_A \Gamma(a_i)^{-1} y^{a_i-1} \exp(- y) \, dy .
\]
Now we define $a_+ := \sum_{i=0}^m a_i$, $G_+ := \sum_{i=0}^m G_i$ and
\[
	\tilde{\bs{B}} \ := \ (G_i/G_+)_{i=0}^m, \quad
	\bs{B} \ := \ (G_i/G_+)_{i=1}^m .
\]
Note that $\tilde{\bs{B}}$ is contained in the unit simplex in $\R^{m+1}$, while $\bs{B}$ is contained in the open set $\mathcal{T}_m = \bigl\{ \bs{u} \in (0,1)^m : u_+ < 1 \bigr\}$ with $u_+ := \sum_{i=1}^m u_i$. We also define $u_0 := 1 - u_+$ for any $\bs{u} \in \mathcal{T}_m$.

\begin{Lemma}
\label{lemma: Gamma-Beta}
The random vector $\bs{B}$ and the random variable $G_+$ are stochastically independent. Moreover,
\[
	G_+ \ \sim \ \mathrm{Gamma}(a_+)
\]
while $\bs{B}$ is distributed according to the Lebesgue density
\[
	f(\bs{u}) \ := \ \frac{\Gamma(a_+)}{\prod_{i=0}^m \Gamma(a_i)} \,
		\prod_{i=0}^m u_i^{a_i-1}
\]
on $\mathcal{T}_m$. For arbitrary numbers $k_0, k_1, \ldots, k_m \ge 0$ and $k_+ := \sum_{i=0}^m k_i$,
\[
	 \Ex \Bigl( \prod_{i=0}^m B_i^{k_i} \Bigr)
	\ = \ \frac{\Gamma(a_+)}{\Gamma(a_+ + k_+)}
		\prod_{i=0}^m \frac{\Gamma(a_i + k_i)}{\Gamma(a_i)} .
\]
\end{Lemma}

As a by-product of this lemma we obtain the following formula:

\begin{Corollary}
\label{cor: Gamma-Beta}
For arbitrary numbers $a_0, a_1, \ldots, a_m > 0$,
\[
	\int_{\mathcal{T}_m} \prod_{i=0}^m u_i^{a_i-1} \, d\bs{u}
	\ = \ \Gamma(a_+)^{-1} \prod_{i=0}^m \Gamma(a_i) .
\]
\end{Corollary}

\begin{proof}[\bf Proof of Lemma~\ref{lemma: Gamma-Beta}]
Note that $\bs{G} = (G_i)_{i=0}^m$ my be written as $\Xi(G_+, \bs{B})$ with the bijective mapping $\Xi : (0,\infty) \times \mathcal{T}_m \to (0,\infty)^{m+1}$,
\[
	\Xi(s, \bs{u}) \ := \ (s u_i)_{i=0}^m .
\]
Note also that
\[
	\det D\Xi(s, \bs{u})
	\ = \ \det \begin{pmatrix}
			u_0    & -s     & -s     & \cdots & -s     \\
			u_1    & s      & 0      & \cdots & 0      \\
			u_2    & 0      & s      & \ddots & \vdots \\
			\vdots & \vdots & \vdots & \vdots & 0      \\
			u_m    & 0      & \cdots & 0      & s      
		\end{pmatrix}
	\ = \ \det \begin{pmatrix}
			1      & 0      & 0      & \cdots & 0      \\
			u_1    & s      & 0      & \cdots & 0      \\
			u_2    & 0      & s      & \ddots & \vdots \\
			\vdots & \vdots & \vdots & \vdots & 0      \\
			u_m    & 0      & \cdots & 0      & s      
		\end{pmatrix}
	\ = \ s^m .
\]
Thus the distribution of $(G_+, \bs{B})$ has a Lebesgue density $h$ on $(0,\infty) \times \mathcal{T}_m$ which is given by
\begin{align*}
	h(s, \bs{u}) \
	&= \ \prod_{i=0}^m
			\bigl( \Gamma(a_i)^{-1} \Xi(s, \bs{u})_i^{a_i-1} \exp(- \Xi(s, \bs{u})_i) \bigr)
		\cdot \bigl| \det D\Xi(s, \bs{u}) \bigr| \\
	&= \ \prod_{i=0}^m
			\bigl( \Gamma(a_i)^{-1} (s u_i)^{a_i-1} \exp(- s u_i) \bigr)
		\cdot s^m \\
	&= \ s^{a_+ - 1} \exp(- s) \prod_{i=0}^m
			\bigl( \Gamma(a_i)^{-1} u_i^{a_i-1} \bigr) \\
	&= \ \Gamma(a_+)^{-1} s^{a_+ - 1} \exp(- s) \cdot f(\bs{u}) .
\end{align*}
Since this is the density of $\mathrm{Gamma}(a_+)$ at $s$ times $f(\bs{u})$, we see that $G_+$ and $\bs{B}$ are stochastically independent, where $G_+$ has distribution $\mathrm{Gamma}(a_+)$, and that $f$ is indeed a probability density on $\mathcal{T}_m$ describing the distribution of $\bs{B}$.

The fact that $f$ integrates to one over $\mathcal{T}_m$ entails Corollary~\ref{cor: Gamma-Beta}. But then we can conclude that
\begin{align*}
	\Ex \Bigl( \prod_{i=0}^m B_i^{k(i)} \Bigr) \
	&= \ \int_{\mathcal{T}_m} \prod_{i=0}^m u_i^{a_i + k_i - 1} \, d\bs{u}
		\Big/ \int_{\mathcal{T}_m} \prod_{i=0}^m u_i^{a_i - 1} \, d\bs{u} \\
	&= \ \frac{\Gamma(a_+)}{\Gamma(a_+ + k_+)}
			\prod_{i=0}^m \frac{\Gamma(a_i + k_i)}{\Gamma(a_i)} .
\end{align*}\\[-5ex]
\end{proof}


\end{document}